# Fully spin-transparent magnetic interfaces enabled by insertion of a thin paramagnetic NiO layer

Lijun Zhu[1,2*], Lujun Zhu[2], and Robert A. Buhrman[1]

1. *Cornell University, Ithaca, New York 14850, USA*
2. *State Key Laboratory of Superlattices and Microstructures, Institute of Semiconductors, Chinese Academy of Sciences, P.O. Box 912, Beijing 100083, China*
3. *College of Physics and Information Technology, Shaanxi Normal University, Xi'an 710062, China*

Spin backflow and spin-memory loss have been well established to considerably lower the interfacial spin transmissivity of metallic magnetic interfaces and thus the energy efficiency of spin-orbit torque technologies. Here we report that spin backflow and spin-memory loss at Pt-based heavy metal/ferromagnet interfaces can be effectively eliminated by inserting an insulating paramagnetic NiO layer of optimum thickness. The latter enables the thermal magnon-mediated essentially unity spin-current transmission at room temperature due to considerably enhanced effective spin-mixing conductance of the interface. As a result, we obtain dampinglike spin-orbit torque efficiency per unit current density of up to 0.8 as detected by the standard technology ferromagnet FeCoB and others, which reaches the expected upper-limit spin Hall ratio of Pt. We establish that Pt/NiO and Pt-Hf/NiO are two energy-efficient, integration-friendly, and high-endurance spin-current generators that provide >100 times greater energy efficiency than sputter-deposited topological insulators BiSb and BiSe. Our finding will benefit spin-orbitronic research and advance spin-torque technologies.

Spin-orbit torques (SOTs)[1-2] have great potential for enabling ultrafast energy-efficient magnetic memories [3,4] and logic [5] for many key electronics applications (e.g. large-scale computing and machine learning). However, energy-efficient, integration-friendly, and high-endurance spin-current generators, the indispensable basis for a successful SOT technology, have remained a major challenge after a decade of intensive exploration. The topological insulators BiSb [6,7] and BiSe [8], despite their high spin Hall ratios ($\theta_{SH}$), are problematic because of their giant resistivities ($\rho_{xx}$)[7,8] and poor thermal and chemical stabilities [9,10]. While some Pt-based heavy metals (HMs) with giant intrinsic spin Hall conductivity ($\sigma_{SH}$) and low $\rho_{xx}$ [11-15] are integration-friendly high-endurance spin current generators, their energy-efficiency is lowered *considerably* by spin backflow (SBF) and sometime also by spin memory loss (SML) at the HM/ferromagnet (FM) interfaces [16-24].

As schematically shown in Fig. 1(a), the spin current generated by the spin Hall effect (SHE) of the HM drops sharply at the metallic HM/FM interface because of the degradation of the interfacial spin transparency ($T_{int}$) by SBF [16-18] and SML [19-24]. In the case of Pt-based HM/FM interfaces, the drift-diffusion analysis [16-18] indicates that SBF reduces $T_{int}$, thus the dampinglike SOT efficiency per applied electric field $\xi_{DL}^E \equiv (2e/\hbar)T_{int}\sigma_{SH}$ and dampinglike SOT efficiency per unit current density $\xi_{DL}^j \equiv T_{int}\theta_{SH}$ by more than a factor of 2 [12,15], while further reduction of $T_{int}$ by SML will occur when the interfacial spin-orbit coupling (ISOC) is significant (e.g. at annealed Pt/Co interfaces)[20,22-24]. Consequently, the optimized Pt-based elemental film [11,12,20], alloys [12], and multilayers [15] that can have giant $\theta_{SH}$ of up to ≈ 0.6-0.8 only provide $\xi_{DL}^j$ of <0.4 ($\xi_{DL}^j$ = 0.16-0.22 for pure Pt [11-13]). While SML can be reduced by interface engineering (e.g. by interface alloying [25] or by insertion of an ultrathin passivating layer [20]), substantial SBF is inevitable at a metallic HM/FM interface when conduction electrons transport the spin angular momentum. So far, there has been no report on unity $T_{int}$, or equivalently $\xi_{DL}^j = \theta_{SH}$, in a Pt-based HM/FM system.

Recently, it has been established that spin current is transmittable in insulating NiO layers [26-31], providing an alternative, electron-free scheme for spin transport. However, both the mechanism and efficiency of the spin transport in NiO have remained in dispute. While some works argue that spin transport in NiO is mediated by coherent antiferromagnetic (AF) magnons [26,28,31] or by tunneling electrons [31], others suggest that the carriers of spin current can only be short-range thermal magnons [27,29]. The insertion of a thin NiO layer between a source and a detector of spin current has also been reported to significantly enhance [26,27,29,30], to abruptly suppress [31], or to have no effect on the spin transmission [28]. Notably, none of the previous experiments [26-31] has evaluated the values of $T_{int}$ of NiO interfaces, and none of the inverse SHE experiments [26,27,29] has discussed the values of $T_{int}\theta_{SH}$ of their yttrium iron garnet (YIG)/NiO/HM samples. For SOT technologies [3-5], exerting a strong SOT on metallic FeCoB is more relevant and more important than increasing the inverse SHE voltage of insulating YIG/HM [26,29]. So far, the only report on enhancement of $\xi_{DL}^j$ by NiO insertion is for Pt/CoTb [30], but the optimized $\xi_{DL}^j$ in that work was below 0.09.

In this Letter, from direct SOT studies based on different techniques and material series, we, for the first time, identify that in a SOT process both SBF and SML at a metallic magnetic interface can be effectively eliminated at room temperature by the insertion of a thin paramagnetic NiO layer of optimum thickness ($t_{NiO}$~0.9 nm). The latter enables thermal magnon-mediated essentially unity spin-current transmission from the HM to the FM [Fig. 1(b)]. As a result, we obtained $\xi_{DL}^j$ of up to 0.8, which is the expected upper-limit $\theta_{SH}$ of Pt [15].

For this study, we sputter-deposited two in-plane magnetized sample series: Pt 4/NiO 0-2.7/FeCoB 1.4 and Pt-Hf/NiO 0-2.7/FeCoB 1.4. Here, the numbers are the layer thicknesses in nm, FeCoB = $Fe_{60}Co_{20}B_{20}$, Pt-Hf = [Pt 0.6/Hf 0.2]$_5$/Pt 0.6. More details on the samples and experimental methods can be found in the Supplementary Materials [32]. The NiO layers are insulating ($\rho_{xx}$>10$^7$ μΩ cm), of disordered polycrystalline face-centered cubic structure (Fig. S1 in [32]), and paramagnetic at room temperature. The minimal exchange bias field ($H_{EB}$) and enhanced coercivity ($H_c$) at room temperature (Figs. 2(a)-2(c) and Fig. S9 in [32]) suggest that the Néel temperature ($T_N$) and the blocking temperature ($T_B$) are well below 300 K in the studied $t_{NiO}$ range ($T_N$ is close to and usually only slightly higher than $T_B$ [29]). This is because FM/AF exchange coupling should occur below $T_B$ and lead to non-zero $H_{EB}$ and enhanced $H_c$. As indicated by the temperature dependences of $H_{EB}$ and $H_c$ in Figs. 2(a)-2(c), $T_B$ for our 0.9 nm NiO is ≈125 K, which is lower than 170 K in [29] but much higher than 15 K in [31] for similar $t_{NiO}$.

The dampinglike SOT efficiencies are determined by angle-dependent "in-plane" harmonic Hall response measurement [12,39]. In the macrospin approximation, the second harmonic Hall voltage response ($V_{2\omega}$) to SOTs under an in-plane magnetic field ($H_{in}$) is given by $V_{2\omega} = V_a \cos\varphi + V_p \cos\varphi \cos2\varphi$ [12,39], where $\varphi$ is the angle of $H_{in}$ with respect to the current direction, $V_a = -V_{AH}H_{DL}/2(H_{in}+H_k)+V_{ANE}$, $V_p$ is the contribution from the fieldlike SOT and the Oersted field, $V_{AH}$ is the anomalous Hall voltage, $V_{ANE}$ the anomalous Nernst effect due to the vertical thermal gradient, $H_k$ the perpendicular anisotropy field, and $H_{DL}$ the dampinglike SOT field. We determine the values of $V_{AH}$ and $H_k$ from the dependence of the first harmonic response Hall voltage ($V_{1\omega}$) on the swept out-of-plane field ($H_z$) under zero $H_{in}$ (see Fig. 2(d) and Fig. S2 in [32]). We obtain $V_a$ for each magnitude of $H_{in}$ from the $\varphi$ dependence of $V_{2\omega}$ [Fig. 2(e)]. The slopes and the intercepts of the linear fits of $V_a$ vs $V_{AH}/2(H_{in}+H_k)$ give the values of $H_{DL}$ and $V_{ANE}$, respectively [Fig. 2(f)]. We note that separation of the $V_{ANE}$ term from the $-V_{AH}H_{DL}/2(H_{in}+H_k)$ term by performing $\varphi$-dependent measurement is critical for a correct estimation of SOTs (i.e. $H_{DL}$ and $\xi_{DL}^{E(j)}$)[12,39]. $\xi_{DL}^{E(j)}$ of the samples can be determined following [14]

$$\xi_{DL}^{E} = \mu_0 H_{DL} M_s t_{CoPt}/E, \quad (1)$$
$$\xi_{DL}^{j} = (2e/\hbar) \mu_0 H_{DL} M_s t_{CoPt}/j_c, \quad (2)$$

where the charge current density $j_c = E\sigma_{xx}$. For our measurements, the applied electric field is $E \approx 66.7$ kV/m; the conductivity $\sigma_{xx}$ varies from $2.3 \times 10^6$ to $3.2 \times 10^6$ $\Omega^{-1}$ m$^{-1}$ for the Pt layers and from $0.71 \times 10^6$ to $0.91 \times 10^6$ $\Omega^{-1}$ m$^{-1}$ for the Pt-Hf layer; $M_s \approx$ 1100-1400 emu/cm$^3$ is the effective magnetization of the FeCoB layer that incudes any magnetic dead layer and magnetic proximity effect (Fig. S2 in [32]).

As summarized in Figs. 2(g) and 2(h), in the absence of a NiO insertion layer, $\xi_{DL}^{E} \approx 4.8 \times 10^5$ $\Omega^{-1}$ m$^{-1}$ ($\xi_{DL}^{j} \approx 0.18$) for Pt 4/FeCoB 1.4 and $\xi_{DL}^{E} \approx 3.5 \times 10^5$ $\Omega^{-1}$ m$^{-1}$ ($\xi_{DL}^{j} \approx 0.37$) for Pt-Hf/FeCoB 1.4, which agree with previous reports [11-13,40]. As $t_{NiO}$ increases, $\xi_{DL}^{E}$ for each of the Pt/NiO/FeCoB and Pt-Hf/NiO/FeCoB series first increases rapidly towards a maximum value at $t_{NiO} \approx 0.9$ nm and then drops down to negligibly small value at $t_{NiO} = 2.7$ nm. $\xi_{DL}^{j}$ ($= \xi_{DL}^{E}/\sigma_{xx}$) is also maximized at $t_{NiO} \approx 0.9$ nm because there is only a weak variation of $\sigma_{xx}$ with insertion of the NiO. The threefold (twofold) enhancement of $\xi_{DL}^{E}$ for Pt/NiO/FM (Pt-Hf/NiO/FM) is also confirmed by spin-torque ferromagnetic resonance measurements [32].

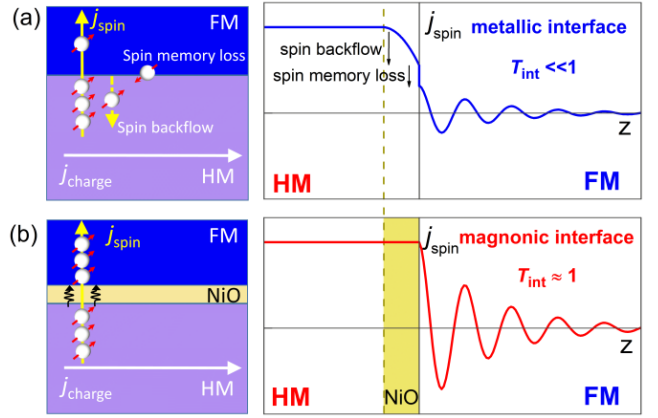

Fig. 1. (a) Metallic HM/FM interface where electron-mediated spin current diffuses from the HM to the FM and undergoes substantial SBF and SML ($T_{int} \ll 1$); (b) Magnonic HM/NiO/FM interface where thermal magnon-mediated spin transport is free of SBF and SML ($T_{int} \approx 1$) at the optimized thickness of the paramagnetic NiO. Here, the spin current flows perpendicular to the layers with spins pointing perpendicular to the magnetization. In the FM the spin current oscillates due to the rapid precession of the spin component that is transverse to the magnetization [16].

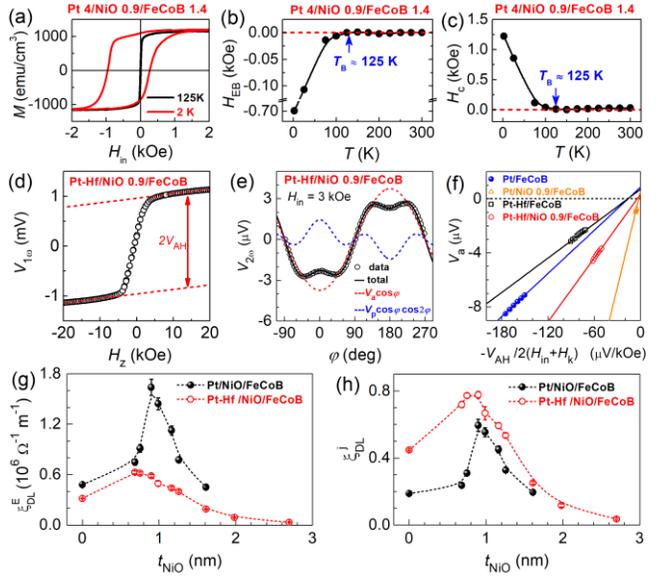

Fig. 2. (a) Magnetization hysteresis of Pt 4/NiO 0.9/FeCoB 1.4 at 2 K and 125 K. Temperature ($T$) dependence of (b) Exchange bias field and (c) Coercivity of the Pt 4/NiO 0.9/FeCoB 1.4, indicating a blocking temperature of ≈ 125 K for the 0.9 nm NiO layer. (d) First harmonic Hall voltage response ($V_{1\omega}$) vs $H_z$, (e) Second harmonic Hall voltage response ($V_{2\omega}$) vs $\varphi$ ($H_{in}$ = 3 kOe), (f) Linear dependence of $V_a$ on $-V_{AH}/2(H_{in}+H_k)$ ($H_{in}$ = 1.5-3.25 kOe), with the slopes being $H_{DL}$, (g) $\xi_{DL}^{E}$ and (h) $\xi_{DL}^{j}$ for the HM/NiO $t_{NiO}$/FeCoB 1.4 (HM = Pt 4 or [Pt 0.6/Hf 0.2]$_5$/Pt 0.6) with different $t_{NiO}$.

This enhancement of the SOTs is not due to any spin current generation from the NiO layers, the NiO interfaces, or bulk effects in the FM since we measured a negligibly small $\xi_{DL}^E$ for the control stack of Ta 1/NiO 0.9/FeCoB 1.4/Hf 0.1/MgO 2/Ta 1.5 (Fig. S4 in [32]). This first strongly suggests that the enhancement of $\xi_{DL}^E$ in the HM/NiO/FeCoB is not due to a recently proposed mechanism [41] whereby a bi-axially anisotropic AF NiO single crystal can magnify (generate) spin current by spin angular momentum influx from the crystal lattice. This is likely because our sputter-deposited thin NiO layers are disordered and paramagnetic. The absence of any important interfacial Rashba-Edelstein effect in these samples is reaffirmed by the small fieldlike/dampinglike torque ratio (<0.1) [18] and the rapid decrease of both torques in the thick NiO limit (Fig. S5 in [32]). The minimal $\xi_{DL}^E$ for another control stack of Pt 4/MgO 1.5/FeCoB suggests negligible spin current generation at the Pt/oxide interfaces of our samples. Therefore, the bulk SHE of Pt and Pt-Hf multilayers is the only important source for the dampinglike SOT in the Pt/NiO/FeCoB and Pt-Hf/NiO/FeCoB. As we discuss below, the dramatic evolution of $\xi_{DL}^E$ with $t_{NiO}$ for the HM/NiO/FM trilayer [Fig. 2(e)] is attributed to first the increase of $T_{int}$ with $t_{NiO}$ to essentially unity by the elimination of SBF, and then to a quasi-exponential decrease of $T_{int}$ due to increasing spin attenuation within the NiO layer as $t_{NiO}$ is beyond its optimal value.

It has been established that the dampinglike SOT in Pt-based alloys [12] and multilayers [15,40] is predominantly from the intrinsic SHE, with the signature being the characteristic reduction of $\sigma_{SH}$ with carrier lifetime (or $\sigma_{xx}$) in the dirty limit [42]. Figure 3(a) shows the $\sigma_{xx}$ dependence of the measured apparent spin Hall conductivity $T_{int}\sigma_{SH} = (\hbar/2e)\xi_{DL}^E$ and the estimated internal spin Hall conductivity $\sigma_{SH} = (\hbar/2e)\xi_{DL}^E/T_{int}$ for three different series of Pt-based materials: Pt-MgO alloys [12], Pt-Ti multilayers [15], and Pt-Hf multilayers [40]. In each case, both $T_{int}\sigma_{SH}$ and $\sigma_{SH}$ decrease rapidly with decreasing $\sigma_{xx}$ as expected for the intrinsic SHE in the dirty limit ($\sigma_{xx} < 4 \times 10^6$ $\Omega^{-1}$-m$^{-1}$)[42]. Our determination of $T_{int}$ is discussed in detail in [32]. For these materials, SML is sufficiently weak so that the drift-diffusion analysis is approximately independent of SML ($T_{int}^{SML} > 0.9$ as determined from its linear dependence on ISOC [20]).

The rapid *increase* of $\xi_{DL}^E$ in the region of $t_{NiO} < 0.9$ nm together with the unity $T_{int}$ at $t_{NiO} \approx 0.9$ safely excludes the possibility of angular momentum transfer via electron tunneling through the insulating NiO layer. The latter, if important, would lead to a monotonic rapid decrease of $\xi_{DL}^E$ with increasing $t_{NiO}$ [26,27,29-31]. It has also been consistently found that insertion of a thin non-magnetic insulator layer (e.g. SrTiO$_3$ [26], AlO$_x$ [29], SiO$_2$ [26,28], or MgO [30]) would degrade rather than enhance spin transmission, with a typical spin attenuation length ($\lambda_s$) of < 0.25 nm. Coherent AF magnons are apparently absent at room temperature in our HM/NiO/FM systems where the NiO is paramagnetic. This is reaffirmed by the very short $\lambda_s$ of < 1 nm as indicated by the rapid, seemly exponential, decrease of $\xi_{DL}^E$ ($T_{int}$) as $t_{NiO}$ is increased above 0.9 nm. In sharp contrast, $\lambda_s$ is very long (e.g. 5 nm [26,30] or >30 nm [31]) for coherent AF magnons in NiO layers that were prepared with different protocols and have large thicknesses and well-ordered crystal structures [26,28]. The *dc* spin current transmission in our paramagnetic NiO samples should be irreverent to coherent evanescent GHz spin waves that was argued to mediate *ac* spin current through epitaxial AF NiO (001)[43]. Therefore, short-range thermal magnons [29,44] are left as the only possible carriers for the highly efficient spin transport through our paramagnetic NiO. Note that the thermal magnons whose wavelength is shorter than the short-range spin correlation of the NiO remains above the $T_N$ [29]. We find that this critical role of thermal magnons is suppressed at low temperatures where AF ordering of the NiO becomes increasingly restored (Section 9 in [32]), in consistence with the expectation that a large magnon band gap prohibits the excitation of thermal magnons and transmission of low-energy spin current [30,44].

Quantitatively, the absence of SBF at the optimal thickness of $t_{NiO} \approx 0.9$ nm suggest that magnonic spin-mixing conductance $G_{eff,m}^{\uparrow\downarrow}$ of the HM/NiO/FM composite interface is comparable to the Sharvin conductance of Pt ($G_{Sh} = 0.68 \times 10^{15}$ $\Omega^{-1}$ m$^{-2}$)[45], the upper bound of the effective spin-mixing conductance of a Pt interface [16,17], so that $T_{int} \approx 2 G_{eff,m}^{\uparrow\downarrow}/G_{HM}$ reaches its limit of 1 (here the spin conductance ($G_{HM}$) is $\approx 1.3 \times 10^{15}$ $\Omega^{-1}$ m$^{-2}$ for Pt [46,47]). It is highly likely that the insertion of the thin insulating paramagnetic NiO blocks the less-efficient electron-mediated spin transport and enables short-range thermal magnon-mediated spin transport with a greatly enhanced $G_{eff,m}^{\uparrow\downarrow}$ [29,44]. Theoretical calculation of $G_{eff,m}^{\uparrow\downarrow}$ directly from the electronic, magnetic, and magnonic properties of the HM/NiO/FM system should be very informative, but is beyond the scope of this Letter.

Note that such impressive enhancement of $T_{int}$ and SOTs we report in the Pt/NiO/FM and Pt-Hf/NiO/FM samples may not be necessarily expected for some other HM/NiO/FM systems. Apparently, $T_{int}$ of those HM/FM samples with $G_{HM}/2 \leq G_{eff,e}^{\uparrow\downarrow}$ already reaches the limit of 1 and thus cannot be enhanced further by any NiO insertion. A good example is the Bi$_2$Se$_3$/NiO/Ni$_{81}$Fe$_{19}$ system [31] where Bi$_2$Se$_3$ has a very low spin conductance ($G_{BiSe} \approx 0.02 \times 10^{15}$ $\Omega^{-1}$ m$^{-2}$ and $G_{eff,e}^{\uparrow\downarrow} \approx 0.60 \times 10^{15}$ $\Omega^{-1}$ m$^{-2}$)[48]. More generally, $G_{eff,m}^{\uparrow\downarrow}$ should vary with $t_{NiO}$ as well as with the types of the HM and the FM because it is determined collectively by the whole "composite" HM/NiO/FM interface rather than solely by the NiO layer. This conclusion is supported by a previous spin Seebeck/inverse SHE experiment [29] that the enhancement of the spin transmission at YIG/NiO 1/HM compared to that of YIG/HM is strong when the HM is Pt but minimal when the HM is Pd or W. Furthermore, we find that $G_{eff,m}^{\uparrow\downarrow}$ at a FM/NiO/Pt (or Pt-Hf) interface can be reduced to below $G_{eff,e}^{\uparrow\downarrow}$ of the corresponding FM/Pt (or Pt-Hf) interface when the FM surface is oxidized and becomes a magnetically dead insulating layer that attenuates spin current (Section 8 in [32]). In addition, insertion of a very thick NiO layer that is AF at room temperature is not beneficial for $T_{int}$ and $\xi_{DL}^{E(j)}$ [28,31] because the AF ordering will suppress excitation of thermal magnons and prohibit the transmission of low-energy spin current [30,44]. For example, insertion of a 25 nm AF NiO layer at $Bi_2Se_3/Ni_{81}Fe_{19}$ interface reduced $\xi_{DL}^j$ from 0.67 to 0.3 [31], the latter is even smaller than that provided by some low-$\rho_{xx}$ metals (e.g. Pt-Hf [40] and Pt-Ti multilayers [15]).

We also note that at interfaces of insulating YIG, where $G_{eff,HM/YIG}^{\uparrow\downarrow}$ mediated only by thermal magnons in YIG [29] can be several times lower than $G_{eff,e}^{\uparrow\downarrow}$ of metallic Pt/FM interfaces [46], a more than a factor of 3 *relative* increase of spin transmission by introducing the enhanced magnons of NiO is possible at the optimal temperatures [29,44]. While the lack of the $T_{int}\theta_{SH}$ values in previous YIG reports [26,27,29,44] prevents evaluation of the exact spin transparency of those YIG/NiO/HM, there is no doubt that the maximum $T_{int}$ of any magnetic interfaces, in either a SOT process or an inverse SHE process, can never exceed unity and thus cannot be greater than that of our Pt (Pt-Hf)/NiO 0.9/FeCoB interfaces.

From the viewpoint of SOT technology, the fact that inserting a thin paramagnetic NiO of optimum thickness between a HM and a FM can result in effectively spin-transparent interfaces for spin transport from the HM to the FM is a very encouraging development. As shown in Fig. 2(f), with the insertion of 0.9 nm NiO layer, $\xi_{DL}^j$ reaches 0.6 for 4 nm Pt ($d$ = 4 nm, $\rho_{xx}$= 37 μΩ cm) and 0.8 for Pt-Hf multilayers ($d$ = 4.6 nm, $\rho_{xx}$=132 μΩ cm). As compared in Figs. 3(b) and 3(c), this results in a SOT device energy efficiency (Section 10 and Table S2 in [32]) of >100 times higher than can be achieved with sputter-deposited β-W ($d$ = 4 nm, $\rho_{xx}$= 300 μΩ cm, $\xi_{DL}^j$= 0.3)[49], BiSb ($d$ = 10 nm, $\rho_{xx}$=1000 μΩ cm, $\xi_{DL}^j$= 1.2)[7], BiSe ($d$ = 4 nm, $\rho_{xx}$=13000 μΩ cm, $\xi_{DL}^j$= 18.6; $d$ = 8 nm, $\rho_{xx}$ = 2150 μΩ cm, $\xi_{DL}^j$= 2.88)[8]. [Note that a recent current-induced coercivity change measurement [6], which is a technique distinctly different from direct harmonic Hall response measurement, reported different results for *single-crystalline* BiSb ($d$ = 10 nm, $\rho_{xx}$ = 400 μΩ cm, $\xi_{DL}^j$= 52). While this would indicate a very low power consumption ($6×10^{-5}$ times of that for the W device [49]), the single-crystalline BiSb requires molecular-beam epitaxy growth on single-crystalline GaAs(100)/MnGa(100)[6], making it disadvantageous for a practical technology that requires integration with CMOS circuits].

In addition to the power efficiency, the Pt/NiO/FM and Pt-Hf/NiO/FM systems are very promising for practical SOT technologies because their low resistivities are also critical for endurance and because of their CMOS integration-friendly properties. The latter includes thermal and chemical stability [20], compatibility with standard sputtering deposition on $SiO_2$ substrate, and ease of being combined with standard high-performance FeCoB magnetic tunnel junctions [4,40,50]. In contrast, BiSb and BiSe, despite their attractive values of $\xi_{DL}^j$ [6-8], suffer from giant resistivities [7,8], thermal instabilities at even moderate temperatures (BiSb melts at 275 °C [9], BiSe sublimates at <280 °C [10]), and chemical instabilities in ambient atmosphere [9,10].

These aspects raise serious questions regarding both the endurance and power efficiency of any SOT devices based on these materials, and pose possibly insurmountable challenges for their successful integration with magnetic tunnel junctions and CMOS circuits. In contrast, Pt/NiO and Pt-Hf/NiO are two exceptionally impressive energy-efficient, integration-friendly, and high-endurance spin-current generators that should immediately benefit the development of practical SOT technologies and further stimulate spin-orbitronic research.

In conclusion, we have presented that SBF and SML at the Pt-based HM/FM interfaces, which considerably degrade the efficiencies of interfacial spin transport and dampinglike SOT, can be effectively eliminated by insertion of an insulating paramagnetic NiO layer of optimum thickness. We find that thermal magnons most likely mediate spin current in the HM/NiO/FM systems and considerably enhance the effective spin-mixing conductance ($G_{eff,m}^{\uparrow\downarrow} >> G_{eff,e}^{\uparrow\downarrow}$). The absence of SML is attributed to the negligible ISOC at the NiO/FM interface. We establish Pt/NiO ($\rho_{xx}$= 37 μΩ cm, $\xi_{DL}^j$=0.6) and Pt-Hf/NiO ($\rho_{xx}$= 132 μΩ cm, $\xi_{DL}^j$= 0.8) as two energy-efficient, integration-friendly, and high-endurance spin-current generators that provide >100 times greater energy efficiency than sputter-deposited BiSb and BiSe. Our finding will immediately benefit spin-orbitronic research and advance SOT technologies.

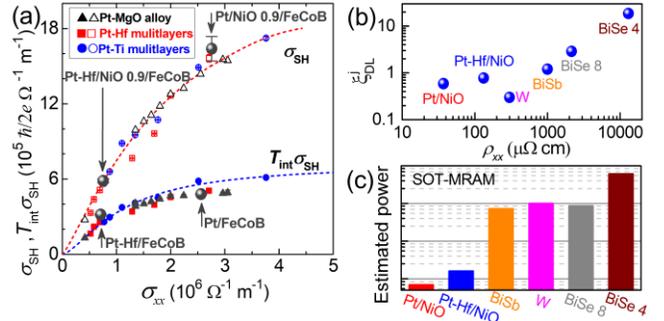

Fig. 3. (a) The variation of the internal ($\sigma_{SH}$) and apparent ($T_{int}\sigma_{SH}$) spin Hall conductivities with electrical conductivity of Pt-based systems. The solid (open) triangle, square, and circles represent $T_{int}\sigma_{SH}$ ($\sigma_{SH}$) of Pt-MgO alloys [12], Pt-Hf multilayers [40], and Pt-Ti multilayers [15], respectively. As indicated by the gray arrows, the four large gray solid dots represent the $T_{int}\sigma_{SH}$ values directly measured for Pt-Hf/FeCoB, Pt-Hf/NiO 0.9/FeCoB, Pt/FeCoB, and Pt /NiO 0.9/FeCoB. While $T_{int}\sigma_{SH}$ for Pt/FeCoB and Pt-Hf/FeCoB overlaps with that for the electron-mediated Pt-MgO/Co, Pt-Hf/Co, and Pt-Ti/Co, $T_{int}\sigma_{SH}$ for the Pt/NiO 0.9/FeCoB and Pt-Hf/NiO 0.9/FeCoB matches the internal bulk values $\sigma_{SH}$, highlighting full spin transmission from the HM to FeCoB in the HM/NiO/FeCoB samples. The dashed lines are for guidance of eyes. (b) $\xi_{DL}^j$ vs $\rho_{xx}$ and (c) Estimated power for SOT-MRAM devices for the sputter-deposited Pt/NiO, Pt-Hf/NiO, BiSb [7], W [49], BiSe 8 nm, and BiSe 4 nm [8]. The power is normalized using that for the W device as unity.

The authors thank R. C. Tapping for instruction on oxide sputtering and Weiwei Lin for fruitful discussions on the spin transport in NiO. The authors acknowledge D. C. Ralph for a critical reading of the manuscript and very helpful suggestions. This work was supported in part by the Office of Naval Research (N00014-15-1-2449), in part by the NSF MRSEC program (DMR-1719875) through the Cornell Center for Materials Research, and in part by the NSF (ECCS-1542081) through use of the Cornell Nanofabrication Facility/National Nanotechnology Coordinated Infrastructure. The TEM measurements performed at Shaanxi Normal University were supported by the National Natural Science Foundation of China (Grant No. 51901121), the Science and Technology Program of Shaanxi Province (Grant No. 2019JQ-433), and the Fundamental Research Funds for the Central Universities (Grant No. GK201903024).

*lz442@cornell.edu

**Supplementary Materials for**

**Fully Spin-transparent magnetic interfaces enabled by insertion of a paramagnetic NiO layer**

Lijun Zhu[1,2*], Lujun Zhu[2], and Robert A. Buhrman[1]

1. Cornell University, Ithaca, New York 14850, USA
2. State Key Laboratory of Superlattices and Microstructures, Institute of Semiconductors, Chinese Academy of Sciences, P.O. Box 912, Beijing 100083, China
3. College of Physics and Information Technology, Shaanxi Normal University, Xi'an 710062, China

* lz442@cornell.edu

**Section 1. Sample preparation and characterizations**

  The samples were deposited at room temperature by sputtering onto Si/SiO$_2$ substrates. The NiO was dc-sputter-deposited using a Ni target at 50 W, with 5 mTorr of 15% O$_2$ + 85% Ar atmosphere. The other layers were deposited with an argon pressure of 2 mTorr. The base pressure was ~10$^{-8}$ Torr. Each of the FeCoB-based samples was capped by a Hf 0.1/MgO 2 nm/Ta 2 nm trilayer that was fully oxidized upon exposure to atmosphere. The structure was characterized by cross-sectional annular dark field scanning transmission electron microscopy (ADF-STEM) and selected area electron diffraction (SAED). The electron beam size is 2 nm in diameter and smaller than the grain size of Pt. Figure S1(a) shows the ADF-STEM image of the Pt 4/NiO 4/NiO 0.9/FeCoB 1.4 stack, indicating the layered structure and of the stack. The Pt layer is polycrystalline, with each grain showing highly ordered lattice periodicity. Figure S1(b) shows the SAED pattern of the Pt layer marked as region ① in Fig. S1(a). The very bright diffraction spots indicate highly ordered face-centered cubic (fcc) (111) structure of the Pt grain. In addition to the Pt fcc (111) spots, the SAED pattern of the Pt/NiO interface [Fig. S1(c)] shows very weak NiO (111) and NiO (110) spots, indicating that the NiO has a polycrystalline, disordered, fcc structure.

  The samples were patterned by photolithography and ion milling into two types of device structures: 5×60 μm$^2$ Hall bars and 10×20 μm$^2$ microstrips, followed by deposition of 5 nm Ti and 150 nm Pt as electrical contacts. The saturation magnetization of each sample was measured with a standard vibrating sample magnetometer (sensitivity ~10$^{-7}$ emu) embedded in a Quantum Design physical property measurement system. During the harmonic response measurements, a Signal Recovery DSP Lock-in Amplifier (Model 7625) was used to source a sinusoidal electric field of $E \approx 66.7$ kV/m onto the Hall bars and to detect the first and second harmonic Hall voltage responses. For the spin-torque ferromagnetic resonance (ST-FMR) measurements, a radio frequency signal generator and a Signal Recovery DSP Lock-in Amplifier (Model 7625) was used and an in-plane magnetic field was swept at 45° with respect to the magnetic microstrip. The conductivity of the heavy metal layers are determined by measuring the conductance enhancement of the stack relative to the control stack with no Pt or Pt-Hf layers (i.e., Ta 1/NiO 0-2.7/FeCoB 1.4/Hf 0.1/MgO 2/Ta 1.5). All the measurements were performed at room temperature.

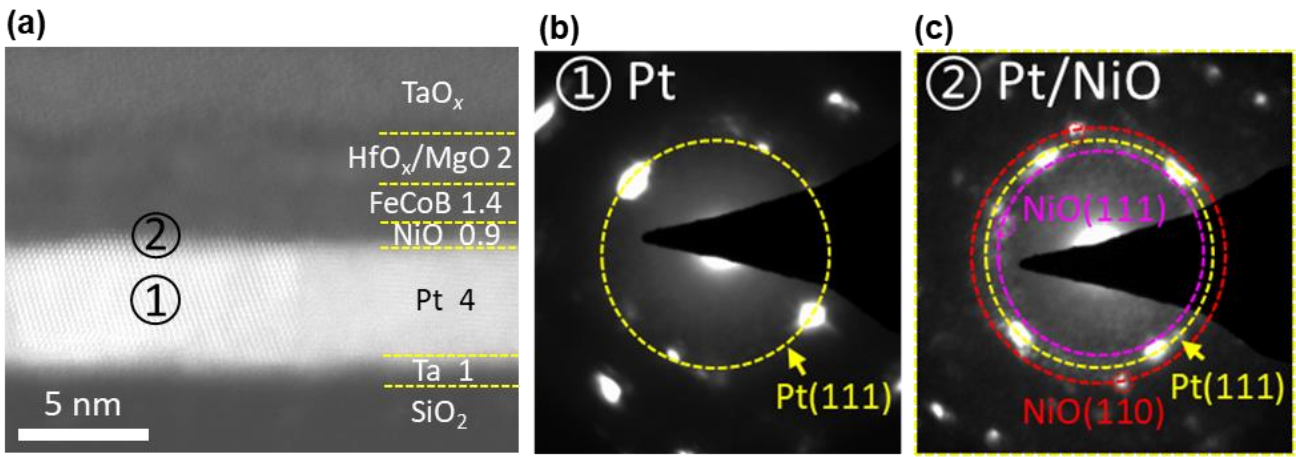

Fig. S1. (a) Cross-sectional annular dark field scanning transmission electron microscopy (ADF-STEM) image of the Pt 4/NiO 4/NiO 0.9/FeCoB 1.4 sample and selected area electron diffraction (SAED) pattern of (b) the Pt (region ① in (a)) and (c) the Pt/NiO interface (region ② in (a)). The small circles indicate dark spots of NiO (110) and NiO (111).

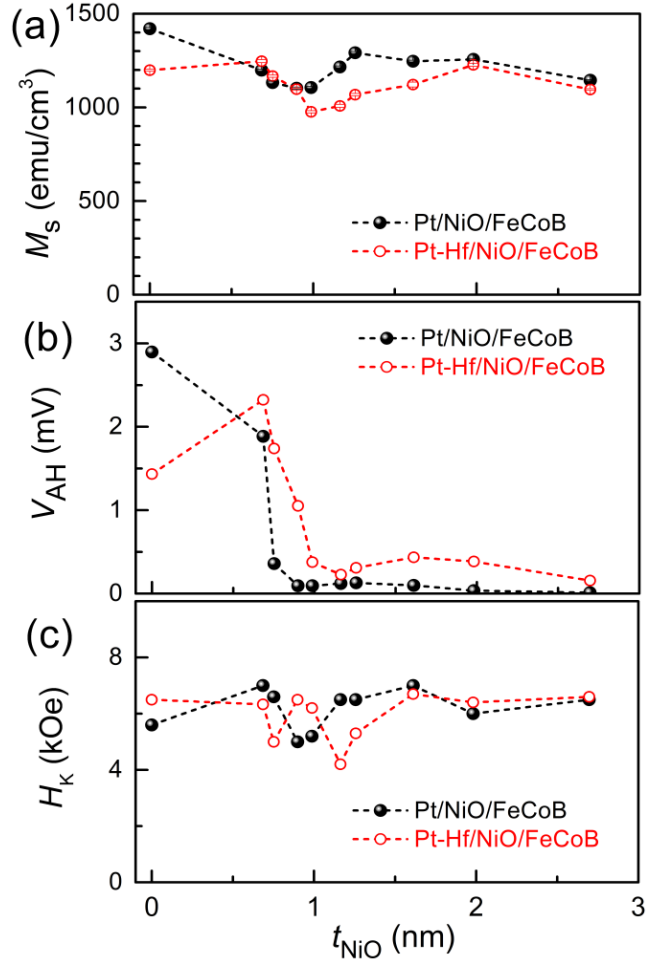

Figure S2. Sample characterizations. (a) Effective saturation magnetization ($M_s$), (b) Anomalous Hall voltage ($V_{AH}$), and (c) Effective magnetic anisotropy field ($H_k$) of the FeCoB layers in the Pt/NiO $t_{NiO}$/FeCoB 1.4 and Pt-Hf/NiO $t_{NiO}$/FeCoB 1.4 samples. $M_s$ was determined by vibrating sample magnetometry, and $V_{AH}$ and $H_k$ were measured using a superconducting physical property measurement system. In (b), the bias electric field is 66.7 kV/m.

## Section 2. Spin-torque ferromagnetic resonance measurement of effect of NiO insertion

As an independent check of the enhancement of the dampinglike SOT efficiency by insertion of a 0.9 nm NiO, we performed FM *thickness-dependent* spin-torque ferromagnetic resonance (ST-FMR) measurement on a bilayer sample Pt 4/Co $t_{Co}$ and a trilayer sample Pt 4/NiO 0.9/Co $t_{Co}$. For the ST-FMR determination, we employ the standard macrospin analysis and assume a negligible spin pumping effect. We first define an effective FMR spin-torque efficiency $\xi_{FMR}$ from the ratio of the symmetric ($S$) and anti-symmetric ($A$) components of the magnetoresistance response of the ST-FMR resonance. $S$ is proportional to $H_{DL}$ and $A$ is due to the sum of the Oersted field and the fieldlike SOT effective field. The dampinglike and fieldlike SOT efficiencies per applied electric field, $\xi_{DL}^{E}$ and $\xi_{FL}^{E}$, can then be obtained from the linear dependence of $\xi_{FMR}^{-1}$ on $t_{FM}^{-1}$ when $\xi_{DL}^{E}$, $\xi_{FL}^{E}$, the HM resistivity ($\rho_{HM}$), and $M_s$ are approximately constant over the studied $t_{FM}$ regime [1],

$$\frac{1}{\xi_{FMR}} = \frac{1}{\xi_{DL}^{E}\rho_{HM}}\left(1 + \frac{\hbar}{e}\frac{\xi_{FL}^{E}\rho_{HM}}{\mu_0 M_s d}\frac{1}{t_{FM}}\right). \quad (1)$$

Using $\rho_{xx} = 43$ μΩ cm for the 4 nm Pt layer, $\xi_{DL}^E$ was estimated from the ST-FMR measurement to be $1.3 \times 10^5$ $\Omega^{-1}$ m$^{-1}$ for Pt 4/Co $t_{Co}$ and $3.4 \times 10^5$ $\Omega^{-1}$ m$^{-1}$ for Pt 4/NiO 0.9/Co $t_{Co}$, reaffirming the almost threefold enhancement of the dampinglike SOT efficiency with the insertion of the 0.9 nm NiO layer. The $\xi_{DL}^E$ values from the ST-FMR measurement are smaller than those from harmonic response measurements, as has been widely observed for many HM/FM systems [2]. For the Pt 4/FeCoB sample we obtained $\xi_{DL}^E = 1.6 \times 10^5$ $\Omega^{-1}$ m$^{-1}$ from the thickness dependent ST-FMR measurements, which is one third of that from the harmonic response measurement ($\xi_{DL}^E \approx 4.8 \times 10^5$ $\Omega^{-1}$ m$^{-1}$, see the main text). While the physical mechanism for the different results of the two techniques has remained unclear, we do note that the angle-dependent harmonic response measurements, if performed carefully, yield results of $\xi_{DL}^E$ that are consistent with that obtained from antidamping SOT switching of in-plane magnetized 3-terminal magnetic tunnel junctions (e.g. $\xi_{DL}^j = \xi_{DL}^E \rho_{xx} \approx 0.3$ for Au$_{0.25}$Pt$_{0.75}$/FM [3]).

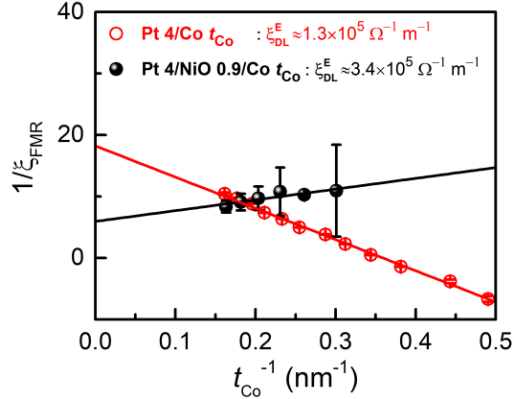

Figure S3. Spin-torque ferromagnetic resonance measurement. Inverse FMR efficiency versus inverse thickness of the Co layer for Pt 4/Co $t_{Co}$ and Pt 4/NiO 0.9/Co $t_{Co}$.

## Section 3. Absence of significant spin-orbit torques from the interfaces

To determine if there was a significant SOT component arising from the NiO/FeCoB or the FeCoB/MgO interfaces we also produced a multilayer consisting of Ta 1/NiO 0.9/FeCoB 1.4/Hf 0.1/MgO, that is a multilayer without either of the heavy metal components. As shown in Fig. S4(a), there was only a minimal second harmonic response from measurements on this sample, indicative of a very small antidamping torque from the spin Hall effect of the very thin Ta (which may be partially oxidized, see Section 8), from the NiO layer, and from interfacial effects[4,5] (Fig. S4(b)). We also performed ST-FMR measurements on this sample and observed a small resonance signal that is predominately anti-symmetric (Fig. S4(c)) and hence is attributable to small net Oersted field, or possibly a fieldlike torque, arising from variation in electron scattering at the NiO/FeCoB and FeCoB/MgO interfaces. In sharp contrast, ST-FMR on a Pt-Hf/NiO 0.9/FeCoB 1.4/Hf 0.1/MgO sample yields a much larger signal (Fig. S4(d)) that is predominately symmetric, indicative of a strong dampinglike spin torque.

Figure S5 compares the harmonic response results of the dampinglike and fieldlike torque efficiencies per unit current density of Pt/NiO $t_{NiO}$/FeCoB 1.4 and Pt-Hf/NiO $t_{NiO}$/FeCoB 1.4 samples. Both torques become negligibly small in the thick limit, while the ratio of the fieldlike and dampinglike torque efficiencies remains very small in the studied thickness range, e.g. < 0.1 for $t_{NiO}$ =0.9 nm. These observations are consistent with a dominant source of the spin Hall effect of the HMs and exclude any important Rashba-Edelstein effect from the interfaces. The Rashba-Edelstein effect, if any, should mainly lead to a local fieldlike torque [4,5].

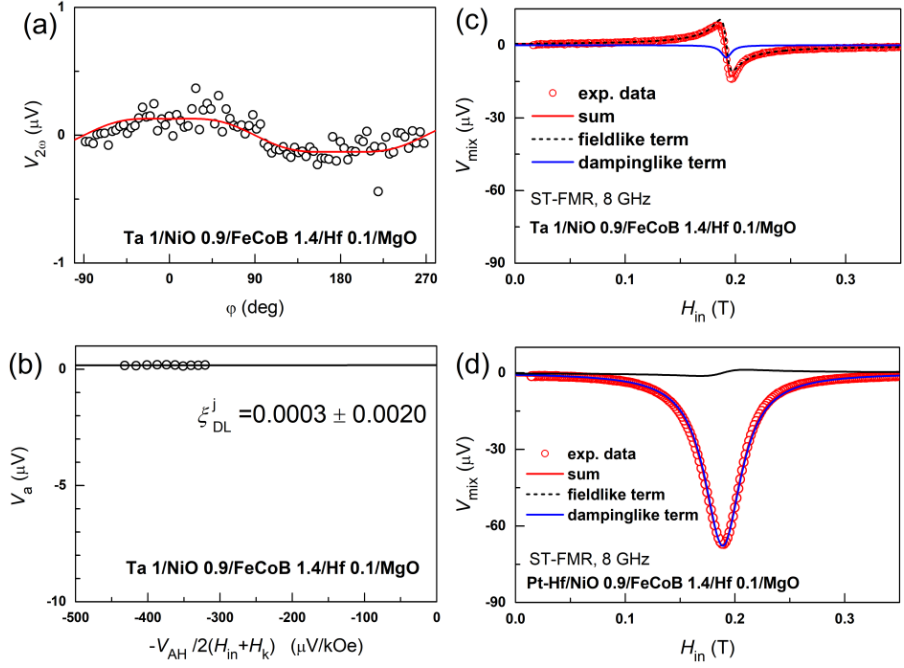

Figure S4. Absence of significant spin-orbit torque contributions from the interfaces of NiO/FeCoB/Hf 0.1/MgO multilayers. a) the second harmonic Hall voltage response ($V_{1\omega}$) for Ta 1/NiO 0.9/FeCoB 1.4/Hf 0.1/MgO as a function of in-plane orientation of magnetization, b) Linear dependence of $V_a$ on $-V_{AH}/2(H_{in}+H_k)$ for Ta 1/NiO 0.9/FeCoB 1.4/MgO, with the slope being the dampinglike effective SOT field. c) Spin-torque ferromagnetic resonance (ST-FMR) spectrum (8 GHz) for Ta 1/NiO 0.9/FeCoB 1.4/Hf 0.1/MgO. d) ST-FMR spectrum (8 GHz) for Pt-Hf/NiO 0.9/FeCoB 1.4/Hf 0.1/MgO.

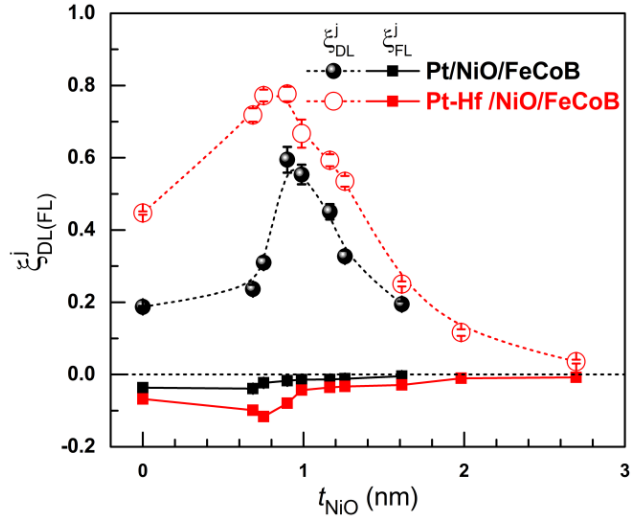

Figure S5. Comparison of the harmonic response results of the dampinglike and fieldlike torque efficiencies per unit current density of Pt/NiO $t_{NiO}$/FeCoB 1.4 and Pt-Hf/NiO $t_{NiO}$/FeCoB 1.4 samples. Both torques become negligibly small in the thick limit, while the ratio of the fieldlike and dampinglike torque efficiencies remains very small in the studied thickness range, e.g. < 0.1 for $t_{NiO}$ =0.9 nm. These results are consistent with that the dominant source of the spin-orbit torques is the spin Hall effect in the HMs and that any contribution from an interfacial Rashba-Edelstein effect is small.

## Section 4. Estimation of interfacial spin transparency

The spin transparency ($T_{int}$) of a heavy metal/ferromagnet (HM/FM) interface is considerably less than unity due to spin backflow (SBF) and spin memory loss (SML). Using a simplifying assumption that the drift-diffusion analysis [4,5] is approximately independent of the weak SML, $T_{int}$ for a SOT process can be estimated as

$$T_{int} = G^{\downarrow\uparrow}_{HM/FM}/(G^{\downarrow\uparrow}_{HM/FM} + G_{SML} + G_{HM}/2), \quad (1)$$

when the thickness ($d_{HM}$) of the HM is much larger than its spin diffusion length ($\lambda_s$). Here $G^{\downarrow\uparrow}_{HM/FM}$ is the bare spin-mixing conductance of the interface, $G_{SML}$ is the effective SML conductance, $G_{HM} = 1/\rho_{xx}\lambda_s$, $\rho_{xx}$, and $\lambda_s$ are the spin conductance, the resistivity, and the spin diffusion length of the HM. Equation (1) can be rewritten as

$$T_{int} = G^{\uparrow\downarrow}_{HM/FM}/(G_{HM}/2 + G^{\uparrow\downarrow}_{HM/FM}) \times [1 + G^{\uparrow\downarrow}_{SML}/(G_{HM}/2 + G^{\uparrow\downarrow}_{HM/FM})]^{-1}, \quad (2)$$

or

$$T_{int} = T^{SBF}_{int} \times T^{SML}_{int}, \quad (3)$$

where $T^{SBF}_{int} = G^{\uparrow\downarrow}_{HM/FM}/(G_{HM}/2 + G^{\uparrow\downarrow}_{HM/FM})$ and $T^{SML}_{int} \equiv [1 + G^{\uparrow\downarrow}_{SML}/(G_{HM}/2 + G^{\uparrow\downarrow}_{HM/FM})]^{-1}$ are the spin transparencies set by the SBF and by the SML, respectively. When $d_{HM}$ of the HM layer is not much greater than its spin diffusion length ($\lambda_s$), the spin transparency set by the SBF should be further modified as [6]

$$T^{SBF}_{int} = [1 - \text{sech}(d_{HM}/\lambda_s)]/[1 + G_{HM}\tanh(d_{HM}/\lambda_s)/2G^{\uparrow\downarrow}_{HM/FM}]. \quad (4)$$

Recently, we have also established that $T^{SML}_{int}$ for a HM/FM interface decreases in a linear manner with increasing interfacial spin-orbit coupling (ISOC) strength [7]. The latter can be measured using the linear indicator of interfacial magnetic anisotropy energy density ($K^{ISOC}_s$) at the HM/FM interface. Specifically, $T^{SML}_{int} \approx 1 - 0.23 K^{ISOC}_s$ for the in-plane magnetized Pt/Co interface with $K^{ISOC}_s$ in erg/cm$^2$.

In the main text, we calculated the internal spin Hall ratio and spin Hall conductivity using the values of $T_{int}$ as calculated from the products of $T^{SBF}_{int}$ and $T^{SML}_{int}$ following Equation (3). $T^{SBF}_{int} \leq 0.5$ for the Pt-based HMs (Pt-MgO alloy, Pt-Ti multilayers, and Pt-Hf multilayers) as estimated using Equation (4), $G^{\uparrow\downarrow}_{HM/FM} \approx 0.59 \times 10^{15}$ $\Omega^{-1}$ m$^{-2}$ (Ref. [5] and [8]), and $G_{HM} \approx 1.3 \times 10^{15}$ $\Omega^{-1}$ m$^{-2}$ (Ref. [6]). As determined from the linear dependence on ISOC [7], SML is a minimal effect for these in-plane magnetized samples ($T^{SML}_{int} \geq 0.9$, see Figure S6 for the example of the Pt-Hf multilayers/Co samples).

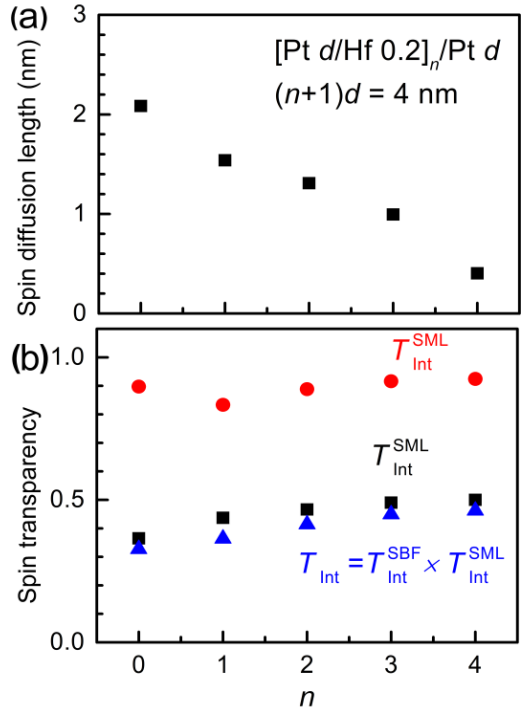

Figure S6. Estimation of the spin transparency. a) The spin diffusion length ($\lambda_s$) for Pt-Hf multilayers. b) The spin transparency for Pt-Hf multilayers/Co with increasing ultrathin Ti insertions. Here Pt-Hf multilayers = [Pt $d$/Hf 0.2]/Pt $d$, with ($m$+1)$d$=4 nm.

**Section 5. Distinct dependences of spin transparency and magnetic damping on the NiO thickness**

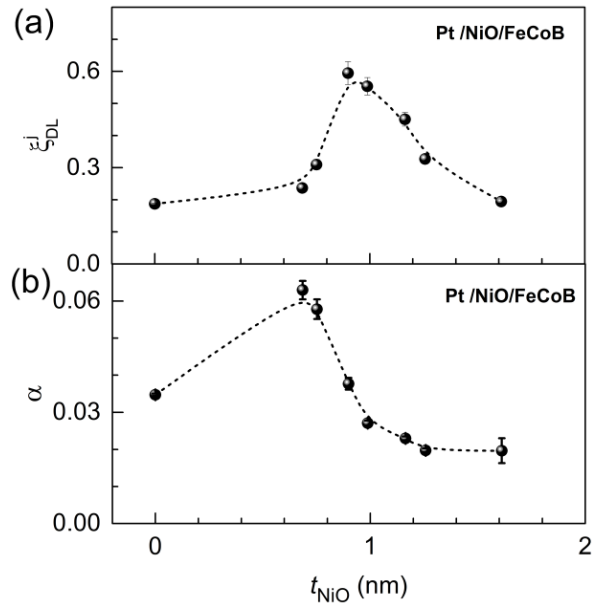

Figure S7. Distinct dependences of a) dampinglike torque efficiency and b) magnetic damping ($\alpha$) of Pt 4/NiO $t_{NiO}$/FeCoB 1.4 on the NiO thickness ($t_{NiO}$), indicating no direct correlation between the interfacial spin transparency and the interfacial roughness. Here $\alpha$ is measured using ST-FMR. In heavy metal/ferromagnet (HM/FM) heterostructures, $\alpha$ is dominated by interfacial non-uniform contribution from the HM/FM interface [8].

**Section 6. Minimal interfacial spin-orbit coupling at NiO/FeCoB interface**

We use the interfacial magnetic anisotropy energy density ($K_s$) as an indicator of the strength of the interfacial spin-orbit coupling (ISOC). Using ST-FMR, we measure the ferromagnetic resonance field ($H_r$) for different thicknesses of the FeCoB as a function of rf frequency ($f$). As shown in Figure S8(a), the effective magnetization ($4\pi M_{eff}$) can be determined by fitting the $H_r$-$f$ data to Kittel's formula [10] $f = (\gamma/2\pi)\sqrt{H_r(H_r + 4\pi M_{eff})}$. In Figure S8(b) we summarize the values of $K_s$ for Pt-Hf/FeCoB 1.4/Hf 0.1/MgO (sample R1), Pt-Hf/NiO 0.9/FeCoB 1.4/Hf 0.1/MgO (sample R2), Ta/FeCoB 1.4/Hf 0.1/MgO (sample R3), as determined following the relation $4\pi M_{eff} \approx 4\pi M_s - 2K_s/M_s t_{FM}$, where $t_{FM}$ is the thickness of the FM layer, and $M_s$ is the saturation magnetization determined by vibrating sample magnetometry (see Figure S2). The total value of $K_s$ is 0.909±0.004 erg/cm$^2$ for Pt-Hf/FeCoB 1.4/Hf 0.1/MgO, 0.892±0.006 erg/cm$^2$ for Pt-Hf/NiO 0.9/FeCoB 1.4/Hf 0.1/MgO, and 0.884±0.005 erg/cm$^2$ for Ta/NiO 0.9/FeCoB 1.4/Hf 0.1/MgO, respectively. After subtracting the contribution of 0.6 erg/cm$^2$ from the top interface, we obtain $K_s$ < 0.31 erg/cm$^2$ for the bottom Pt-Hf/FeCoB and the NiO/FeCoB interface, which is a minimal value compared with those causing a significant spin memory loss in a spin-orbit torque process [7].

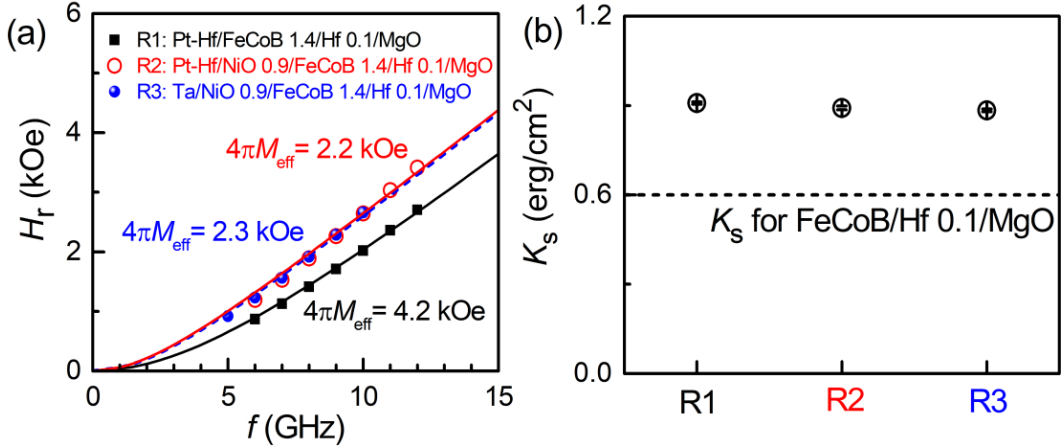

Figure S8. Determination of interfacial spin-orbit coupling at NiO/FeCoB interface. a) Frequency dependence of the FMR resonance field and the determined effective magnetization ($4\pi M_{eff}$) and b) interfacial magnetic anisotropy energy for Pt-Hf/FeCoB 1.4/Hf 0.1/MgO (sample R1), Pt-Hf/NiO 0.9/FeCoB 1.4/Hf 0.1/MgO (sample R2), Ta/FeCoB 1.4/Hf 0.1/MgO (sample R3), respectively. The lines in a) represent fits to Kittel's formula.

**Section 7. Absence of exchange bias at room temperature**

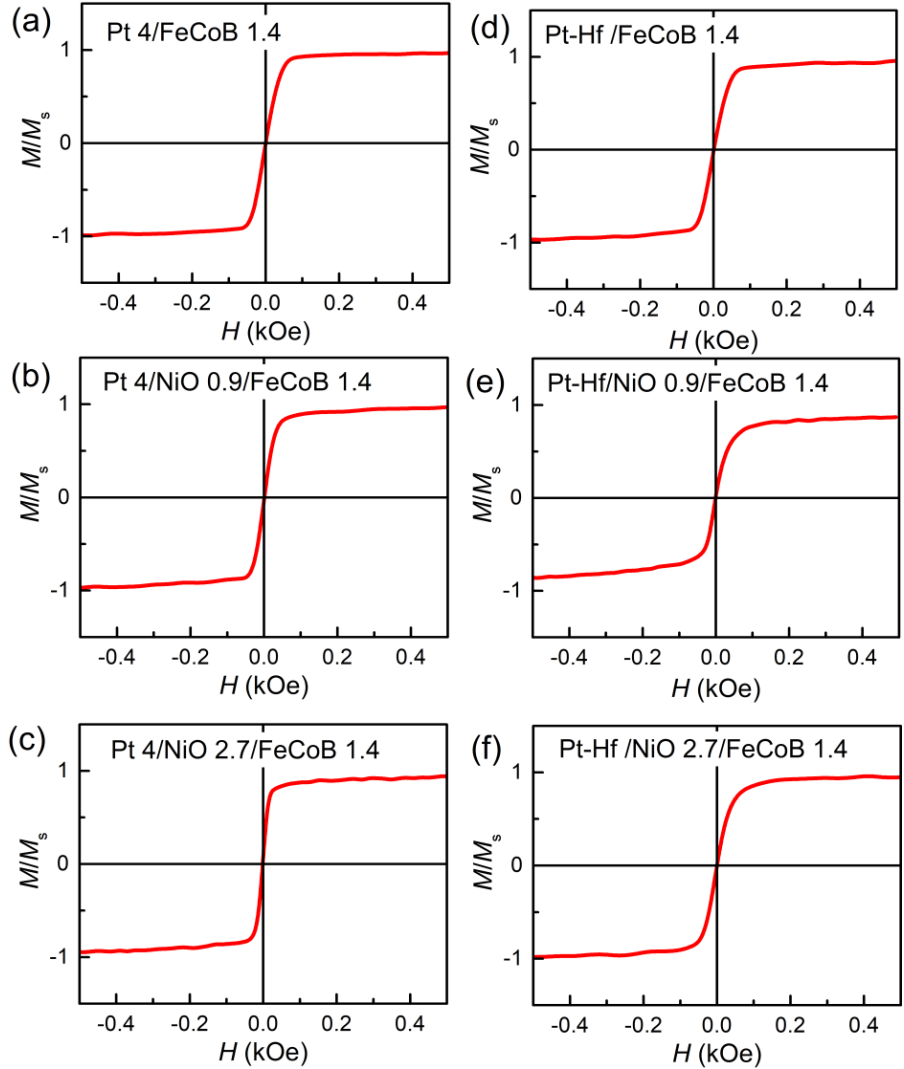

Figure S9. Absence of indication of exchange bias at room temperature. (a) Pt 4/FeCoB 1.4, (b) Pt 4/NiO 0.9/FeCoB 1.4; (c) Pt-Hf/NiO 0.9/FeCoB 1.4; (d) Pt 4/NiO 2.7/FeCoB 1.4; (e) Pt 4/NiO 2.7/FeCoB 1.4; (f) Pt-Hf/NiO 2.7/FeCoB 1.4. In (d)-(f), Pt-Hf=[Pt 0.6/Hf 0.2]$_5$/Pt0.6. While the magnetic field was swept at a slow ramp rate of 1 Oe/sec up to ±2 T for all the measurements, only the data of low field region is shown for clarity. Within the uncertainty, there is no indication of NiO-induced exchange bias. This is consistent with previous reports that the blocking temperature in the sputter-deposited thin NiO is well below 300 K (see Figs. 2(a)-2(c) in the main text and also Refs. [10,12]).

## Section 8. Influence of potential interfacial oxidation

Here we show that the spin transparency of a HM/NiO/FM interface can be reduced to below that of the corresponding HM/FM interface when the first few atomic layers of FM adjacent to the NiO is oxidized and forms a magnetically dead layer. We prepared perpendicularly magnetized sample series of Pt 2/Co 0.6-1.2/NiO 0, 0.9/Pt 4, Pt 2/Co 0.6-1.2/NiO 0, 0.9/Pt-Hf (Pt-Hf = [Pt 0.6/Hf 0.2]$_5$/Pt 0.6), and Pt 2 /Co 0.6-1.2/NiO 0.9 (numbers are layer thicknesses in nm), each of which is seeded by 1 nm Ta layer and capped by a bilayer of MgO 2/Ta 1.5. In Table S1, we calculate the dampinglike spin-torque efficiency contribution of top HM layers ($\xi_{DL,top}^E$) by subtracting that of the bottom 2 nm Pt ($\xi_{DL,bottom}^E$) from the total measured dampinglike torque efficiency ($\xi_{DL,tot}^E$). The Pt 2/Co/NiO 0.9/HM samples show considerably lower effective saturation magnetization ($M_s^{eff}$) of the Co layer compared to the Pt 2/Co/HM samples ($M_s^{eff}$ vary from 1510 to 1610 emu/cm$^3$ due to magnetic proximity effect), indicating that in the stacks with a 0.9 NiO layer the sputtering deposition of the NiO leads to significant oxidation and magnetic dead layer at the top surface of the Co. The level of this oxidation appears to vary from sample to sample depending on the details of deposition, for example, $M_s^{eff}$ is 910 emu/cm$^3$ for Pt 2/Co 0.8/NiO 0.9/Pt 4, 360 emu/cm$^3$ for the Pt 2/Co 1.2/NiO 0.9/Pt-Hf 4, and 560 emu/cm$^3$ for Pt 2/Co 1.2/NiO 0.9. As a consequence, $\xi_{DL,top}^E$ is substantially degraded in the Pt 2/Co/NiO 0.9/HM samples. In the case of Pt 2/Co 1.2/NiO 0.9/Pt-Hf, the torque contribution from the top Pt-Hf layer becomes negligibly small.

This observation suggests that in samples including a nickel oxide layer the possible oxidation of the adjacent layers should be carefully taken care of. Although the Pt and the Pt-terminated Pt-Hf multilayers that we study in the work are robust against oxidation, the surface of many other spin current generators, e.g. W, Ta, Bi$_2$Se$_3$, BiSb, and Bi$_2$Tb$_3$ will likely be substantially oxidized. We note that due to the relatively low electron affinity of Ni, even if the magnonic layer is formed by sputtering from a NiO target it is quite likely that O ions will be liberated in the sputtering process and oxidize the surface of layers with higher electron affinity.

Table S1. Effects of oxidation. $M_s^{eff}$ is the measured effective saturation magnetization ($M_s^{eff}$) of the Co layer by assuming that the measured magnetic moment is contributed by the Co layer with the deposited thickness. The dampinglike torque efficiencies per applied field from the top HM layer ($\xi_{DL,top}^E$) are calculated by subtracting that from the 2 nm bottom Pt layer ($\xi_{DL,bottom}^E$) from the total SOT efficiency ($\xi_{DL,tot}^E$), i.e. $\xi_{DL,top}^E = -\xi_{DL,tot}^E + \xi_{DL,bottom}^E$. Pt-Hf = [Pt 0.6/Hf 0.2]$_5$/Pt 0.6.

| Control samples | $M_s^{eff}$ (emu/cc) | $\xi_{DL,tot}^E$ ($10^5\,\Omega^{-1}\,m^{-1}$) | $\xi_{DL,top}^E$ ($10^5\,\Omega^{-1}\,m^{-1}$) | $\rho_{xx,HM}$ ($\mu\Omega$ cm) |
|---|---|---|---|---|
| Pt 2/Co 0.8/Pt 4 | 1510 | -3.27 | 5.34 | 43.4 |
| Pt 2/Co 0.8/NiO 0.9/Pt 4 | 910 | -2.46 | 4.53 | 59 |
| Pt 2/Co 1.2/Pt-Hf | 1610 | -2.42 | 4.49 | 72 |
| Pt 2/Co 1.2/NiO 0.9/Pt-Hf | 360 | 2.01 | 0.06 | 107 |
| Pt 2 /Co 1.2/NiO 0.9 | 560 | 2.07 | 0 | 83 |

## Section 9. Temperature dependence of the spin-orbit efficiencies in Pt 4/NiO 0.9/FeCoB 1.4

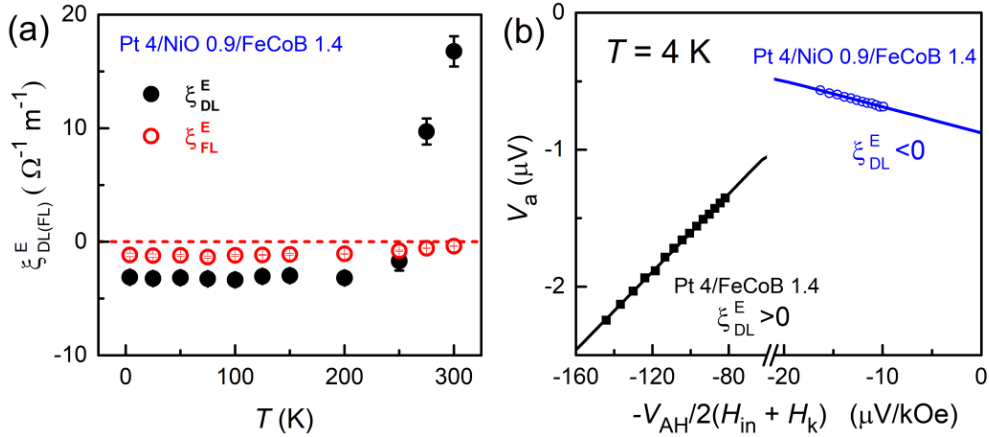

Figure S10. (a) Temperature ($T$) dependence of the dampinglike ($\xi_{DL}^E$) and fieldlike ($\xi_{FL}^E$) spin-orbit efficiencies per applied electric field in Pt 4/NiO 0.9/FeCoB 1.4. (b) Linear dependence of $V_a$ on $-V_{AH}/2(H_{in}+H_k)$ ($H_{in}$ = 2-9 kOe) for Pt 4/NiO 0.9/FeCoB 1.4 and Pt 4/FeCoB 1.4 at 4 K with the slopes being dampinglike spin orbit torque field $H_{DL}$. The slopes indicate a negative $\xi_{DL}^E$ for Pt 4/NiO 0.9/FeCoB 1.4 and a positive $\xi_{DL}^E$ for Pt 4/FeCoB 1.4.

Interestingly, the dampinglike spin-orbit torque (SOT) efficiency per applied electric field, $\xi_{DL}^E$, for Pt 4/NiO 0.9/FeCoB 1.4 drops from $1.7 \times 10^6$ $\Omega^{-1}$ $m^{-1}$ at 300 K to $-0.3 \times 10^6$ $\Omega^{-1}$ $m^{-1}$ at 200 K and then remains roughly constant upon further cooling [Fig. S10(a)]. The rapid reduction of $\xi_{DL}^E$ and thus the spin transparency ($T_{int}$) with temperature is most likely due to the suppressed excitation of thermal magnons as the temperature gradually approaches the Néel temperature ($T_N$) and the blocking temperature ($T_B$) of the NiO (125 K as determined in Fig. 2 in the main text). It is known that the magnon band gap of AF suppresses the excitation of thermal magnons and prohibits low-energy spin current [13].

The negative sign of $\xi_{DL}^E$ for Pt 4/NiO 0.9/FeCoB 1.4 below 250 K is an interesting observation, which is suggestive of additional mechanisms that are likely associated with the antiferromagnetic ordering or paramagnetic-antiferromagnetic transition in the NiO layer. Note that a sign change in the dampinglike SOT is absent in our Pt/FeCoB [Fig. S10(b)] and other HM/FM systems (HM = Pt, W, Ta)[14]. The negative $\xi_{DL}^E$ cannot be attributed to a low-temperature Rashba-Edelstein effect because the fieldlike spin-orbit efficiencies per applied electric field ($\xi_{FL}^E$) remains smaller than $\xi_{DL}^E$ in all the studied temperatures and because the interfacial spin-orbit coupling is negligible at the Pt 4/NiO 0.9/FeCoB 1.4 interface.

An interesting, somewhat surprising, observation is that the spin-orbit torque of our Pt/NiO/FeCoB system does not show any low-temperature peak behavior, in contrast to the spin current transmission in the spin Seebeck/inverse spin Hall experiments of YIG/NiO/Pt systems. We tentatively attribute the different temperature behaviors to the distinct thermal properties of the spin current generators (spin Hall metal Pt in our case, insulating ferromagnet YIG in the case of Ref. [12]), and the physics process (e.g. SOT process should be irrelevant to the thermopower $S$, while in the spin Seebeck case, the spin current injected into the spin detector is proportional to the $S$, with the $S$ shows a peak at the paramagnetic-antiferromagnetic transition). More specifically, in the YIG/NiO/Pt case, the spin current, that is generated by YIG and detected by the Pt, shows a peak at temperatures around the $T_B$ and thus the $T_N$ of the NiO layers (~200 K for 1.2 nm NiO). So far, the existing temperature dependent theories and experiments (e.g. in Ref. [12]) *only* considered the inverse spin Hall process in the YIG/NiO/HM system ——how a spin current generated in YIG transports to HM. However, there has been no report on low-temperature SOTs in a HM/NiO/FM system. A better, detailed understanding of the distinct behaviors of the spin-orbit torque and spin Seebeck effect worth future theoretical and experimental studies.

Here, we do want to note several pieces of important differences between the two experiments. First, the spin current generator Pt in our SOT process has a giant robust SHE (or spin Hall conductivity) that does not decrease or can even increase at low temperatures down to zero K. In contrast, in the case of SSE/ISHE work the thermal spin current generation in the ferromagnetic insulator YIG vanishes at zero K and varies strongly at finite temperatures. We notice that the detected spin current for the SSE process is proportional to the transverse thermopower $S$ that is strongly temperature dependent. $S$ seems to be related to both spin current generation and interfacial spin transmission. In Ref. [12], while both $S$ and the spin current density detected in Pt maximize at around $T_B$ of their NiO layers, *the ratio of the two*, *related to the spin backflow (spin conductance, spin diffusion length, electrical conductivity),* was not enhanced as the temperature decreases from 303 K to well below $T_B$ and thus $T_N$ of the NiO. The spin current detected by the Pt decreases to zero as the temperature approaches 0 K because the lack of thermal magnons in YIG at low temperatures leads to vanishing $S$ in the YIG/NiO/FM system. These characteristics are absent in the SOT process of a Pt/NiO/FM system. In the latter case, the spin-orbit torque is not expected to decrease at 0 K or enhanced at around the $T_N$. At least, $T_{int}$ cannot be enhanced any further upon cooling because it already reaches its upper limit of 1 at 300 K. While spin current should be still transmittable in antiferromagnetic insulator of NiO at low temperatures via assistance of coherent antiferromagnetic magnons that can be excited by spin current, the detailed behaviors and mechanism are yet to study.

We also mention that Ref. [11] reported a small peak at a 25 nm NiO insertion, which could be, at least partly, attributed to the structural improvement and the considerable increased spin attenuation length of the NiO layers. Note that the spin attenuation length was as short as < 0.5 nm when the NiO layer was below 2 nm, and became as long as 30 nm in the thicker region, these would indicates considerable variation of the spin attenuation within the NiO.

## Section 10. Calculation of power consumption of the SOT magnetic tunnel junction devices

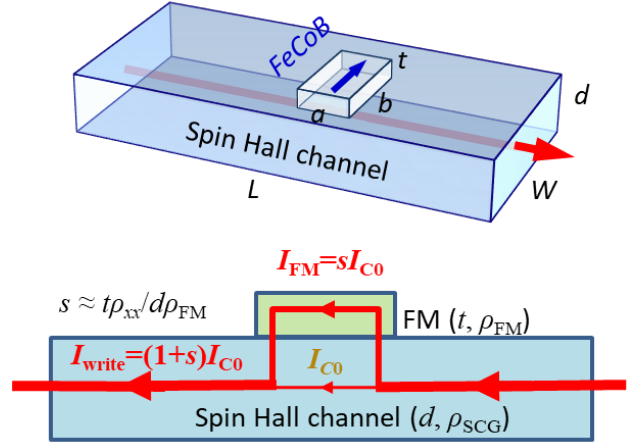

Figure S11. Schematics for a spin-orbit torque MRAM device. (a) Dimension definitions. (b) Current shunting into free layer.

Here we calculate the power efficiency of a model spin-orbit torque MRAM device consisting of a $400\times200\times d$ nm$^3$ spin Hall channel and a $30\times200\times2$ nm$^3$ FeCoB free layer (resistivity $\rho_{FeCoB}$ = 130 μΩ cm) by taking into account the current shunting into the free layer. For convenience of discussion, we note the device dimensions as shown in Fig. S11: channel length $L$ = 400 nm, channel width $W$ = 200 nm, channel thickness $d$; free layer "length" $a$ = 30 nm, free layer "width" $b$ =200 nm, free layer thickness $t$ = 2 nm. Since $t \ll a$ and $t \ll b$, current spreading has a minimal influence and the parallel resistor model can be used. The total write current ($I_{tot}$) in the spin Hall channel is given by

$$I_{tot} = I_{c0}(1+s), \qquad (5)$$

where $I_{c0}$ is the "useful" current in the spin Hall channel that drives the magnetization switching (Fig. S11), $I_{c0}s$ is the "wasted" current shunted into the FeCoB free layer, $s = \rho_{SCG}tb/\rho_{FM}dW$ is the shunting factor, $\rho_{FM}$ and $\rho_{SCG}$ are the resistivities of the free layer and heavy metal layers, respectively. According to the macrospin model, which is found to work reasonably well for in-plane magnetized magnetic tunneling junctions, we have

$$I_{c0} = (2e/\hbar)\mu_0 M_s t_{FM}\alpha(H_c+4\pi M_{eff}/2)/\xi_{DL}^j, \qquad (6)$$

in which $e$, $\mu_0$, $\hbar$, $\alpha$, $H_c$, $M_{eff}$, and $\xi_{DL}$ are the elementary charge, the permeability of vacuum, the reduced Planck constant, the magnetic damping, the coercivity, the effective magnetization, the damping-like spin torque efficiency, respectively. The power consumption is then given by

$$P = I_{c0}^2 [s^2\rho_{FM}\, a/bt + \rho_{SCG}\, a/Wd + (1+s^2)\rho_{SCG}(L-a)/Wd]. \qquad (7)$$

Using Eq. (7), we estimated the power consumption for SOT-MRAM based on different strong spin current generators, the BiSe (4 nm [15], 6 nm [15], and 8 nm [15,16]), BiSb [17], W [18], Au$_{0.25}$Pt$_{0.75}$ [19], Pt-Hf/NiO, and Pt/NiO samples. As summarized in Table 1, the sputter-deposited topological insulators BiSe and BiSb with $\xi_{DL}^j>1$ requires more than 100 times higher power consumption compared to the optimal Pt/NiO we establish in this work. This is because their very high resistivities leads to considerable current shunting into the metallic magnetic layer so that both the total write current and the channel impedance are very high. This result highlights the critical roles of both $\xi_{DL}^j$ and resistivity of the spin current generator. A low resistivity is also indispensable for spin-orbit torque applications that require high endurance (e.g. memories and logic) and low impedance (e.g. cryogenic memories).

We note that the estimated power here is just for the SOT-MRAM itself and does not include the power on the transistors. The latter can be negligible for some cryogenic superconducting transistors, and can also be greater than that of MRAM cell if the transistor resistance is much greater than the channel impedance (see Table S2 for the estimated channel impedance for different spin current generators).

Table S2. Estimated power consumption for SOT-MRAM based on different strong spin current generators. The power is normalized using that of Pt/NiO as the unity. The parameters are that channel length $L$ = 400 nm, channel width $W$ = 200 nm, channel thickness $d$; free layer "length" $a$ = 30 nm, free layer "width" $b$ = 200 nm, free layer thickness $t$ = 2 nm, $\rho_{FeCoB}$ = 130 μΩ cm.

| Spin current generators | Thickness $d$ (nm) | $\xi_{DL}^j$ | Resistivity $\rho_{SCG}$ (μΩ cm) | Shunting factor $s$ | Normalized power | Channel impedence (kΩ) | References |
|---|---|---|---|---|---|---|---|
| BiSe | 4 | 18.6 | 13000 | 50 | 882 | 65 | DC et al. [15] |
| BiSe | 6 | 4.5 | 3100 | 7.9 | 167 | 10.3 | DC et al. [15] |
| BiSe | 8 | 2.88 | 2150 | 4.1 | 125 | 5.4 | DC et al. [15] |
| BiSe | 8 | 3.5 | 1754 | 3.4 | 50 | 4.4 | Millnik et al. [16] |
| BiSb | 10 | 1.2 | 1000 | 1.5 | 104 | 2 | Chi et al. [17] |
| W | 4 | 0.3 | 300 | 1.2 | 145 | 1.5 | Pai et al. [18] |
| Au$_{0.25}$Pt$_{0.75}$ | 5 | 0.35 | 80 | 0.25 | 12 | 0.32 | Zhu et al. [19] |
| Pt-Hf/NiO | 4.6 | 0.8 | 132 | 0 | 2.3 | 0.57 | This work |
| Pt/NiO | 4 | 0.6 | 37 | 0 | 1 | 0.185 | This work |